\documentstyle[twoside,fleqn,espcrc2,epsf]{article}

\newcommand{\beq}{\begin{equation}}

\newcommand{\eeq}{\end{equation}}
\newcommand{\bea}{\begin{eqnarray}}
\newcommand{\eea}{\end{eqnarray}}

\newcommand{\phib}{\bar{\phi}}

\newcommand{\tl}{\tilde{t}_{_L}}
\newcommand{\Ql}{\tilde{Q}_{_L}}
\newcommand{\tr}{\tilde{t}_{_R}}

\newcommand{\postscript}[2]
 {\setlength{\epsfxsize}{#2\vsize}
\setlength{\epsfysize}{#2\vsize}
  \centerline{\epsfbox{#1}}}
\newcommand{\AmS}{{\protect\the\textfont2
  A\kern-.1667em\lower.5ex\hbox{M}\kern-.125emS}}

\title{ Color and charge breaking minima in the MSSM
}

\author{Alexander Kusenko\address{Department of Physics and Astronomy,
University of Pennsylvania, \\
Philadelphia, PA 19104-6396, USA}%
        \thanks{email address: sasha@langacker.hep.upenn.edu; address after
October 1, 1996: Theory Division, CERN, CH-1211 Geneva 23, Switzerland }
}
       
\begin{document}

\begin{abstract}
The scalar potential of theories with broken supersymmetry can have a
number of local minima characterized by different gauge groups.  Symmetry
properties of the physical vacuum  constrain the parameters of the MSSM. 
We discuss  these constraints, in particular those that result from the
vacuum stability with respect to quantum tunneling. 
\end{abstract}

\maketitle

A generic feature of theories with (softly broken) supersymmetry is a
scalar  potential,  $V(\phi)$, that depends on a large number of scalar
fields $\phi=(\phi_1,...,\phi_n)$.   For this reason, the scalar potential
of the  MSSM, unlike that of the Standard Model, may have a number of local
minima characterized by different gauge symmetries.   In particular, the
supersymmetric partners of quarks, $\tilde{Q_{_L}}$ and $\tilde{q_{_R}}$, 
may have non-zero vev in some minima, where the tri-linear terms 
$A H_2 \tilde{Q_{_L}} \tilde{q_{_R}} $  and $\mu H_1 \Ql \tr$  are large and
negative (here $H_{1,2}$ denote Higgs fields and $A$ is the SUSY breaking
parameter).  These color and charge breaking (CCB) minima may be local, or
global, depending on the values of the MSSM parameters.  (Of course, there
might be directions along which the effective potential is unbounded from
below (UFB), in which case all the minima are local.) 

Any of these local minima may serve as the ground state for the Universe at
present, provided that the lifetime of the metastable state
is large in comparison to the age of the Universe.  The latter is plausible
\cite{chh,kls} because the tunneling rate in quantum field theory is
naturally suppressed by the exponential of a typically large dimensionless
number, the saddle point value of the Euclidean action.  

Are there any empirical, or general theoretical considerations that could
rule out the possibility of the Universe at present being in a 
long-lived, but metastable, vacuum?  Apparently, the answer is no. 

Different minima of the scalar potential are characterized by different
values of the cosmological constant.  However, the cosmological constant
problem is just as severe  in the stable vacuum as it is in
a metastable one.  In fact, in a
large class of locally supersymmetric Unified theories the cosmological
constant can be fine-tuned to be zero in any of the {\it local} minima, but 
not in the global minimum \cite{w}.  

In principle, one can imagine a Gedanken experiment to determine whether
the vacuum is true, or false.  However, if a metastable vacuum has existed
for $\tau_{_U}=$ 10 billion years, then any effects of the metastability
\cite{ak3}  would be characterized by the scale $< 1/\tau_{_U} \sim 10^{-33}$
eV,   beyond any hope of being observable.  Sadly, the first direct
evidence of the vacuum instability would be a catastrophic event. 

The question, however, has more than just eschatological relevance.  
It is very important scientifically, because ruling out the possibility that
the Universe rests in the false vacuum would impose strong constraints on
theories of fundamental interactions \cite{chh,kls,ccb}.  Clearly, only the
false vacua whose lifetimes are small in comparison to the age of the
Universe are ruled out empirically.  

\begin{figure}
\postscript{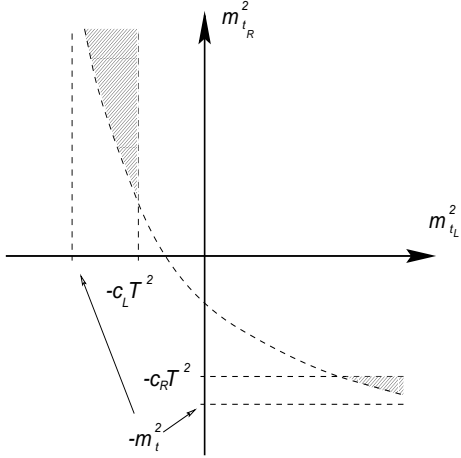}{0.3}
\caption{Region of parameters (shaded) which can be ruled out by requiring
stability of the {$SU(2)\times U(1)$} symmetric minimum above the
electroweak transition temperature. Coefficients {$c_{_L,_R}$} are defined 
in Ref. [2].} 
\label{fig1}
\end{figure}

We concentrate on the TeV-scale CCB minima.  These generally disappear at
temperatures $T \gg 1$ TeV.  Therefore, if the temperature of reheating
after inflation is $10 $ TeV or higher, one can assume that the 
electroweak symmetry is restored.  
Then at $T=T_c \sim 100$ GeV the electroweak phase transition proceeds from
an  $ SU(3)\times SU(2)\times U(1) $ -symmetric phase to the phase of the 
broken symmetry.  It was shown in \cite{kls} that for a wide range of
parameters (Fig. \ref{fig1}) this transition favors the Standard Model-like 
(SML) minimum over the CCB minima. 

The probability of tunneling from the SML into a CCB minimum, 
or a UFB valley, determines the lifetime of a false SML vacuum.  Tunneling
rate can be evaluated using the semiclassical approximation \cite{tunn} and
is proportional to $\exp (-S[\phib])$, where $ S[\phib] $ is the Euclidean
action of the so called ``bounce'', $\phib(x)$, a solution  of the
classical Euclidean field equations.  In practice, however, finding 
$\phib(x)$ numerically is very difficult (or nearly impossible), 
especially in the
case of a potential that depends on more than one scalar field. This is
because $\phib(x)$ is an unstable solution, as it must be to be a saddle
point of the functional $S[\phi]$.  An effective alternative to solving the
equations of motion is to use the method of Ref. \cite{ak1}.  The idea is
to replace the action $S$ with a different functional, $\tilde{S}$, for
which the same solution, $\phib(x)$, is a minimum, rather than a saddle 
point. Then $\phib(x)$ can be found numerically using a
straightforward relaxation technique to minimize $\tilde{S}$.   

Another significant simplification comes from the observation that the
tunneling rate (in semiclassical approximation) is independent 
of the physics at the scales large in comparison to $\phib(0)$, the escape
point.  This non-perturbative decoupling \cite{kls} of high-energy physics
allows one to treat the UFB valleys on the same footing with the very deep
CCB minima. Essentially, one can set a cutoff at, {\it e.\,g.}, $10$ TeV
and, as long as the bounce solutions found numerically do not extend 
beyond this limit, one can justifiably ignore the physics at the higher 
energy scales.

For a 10 billion year old Universe, the tunneling probability is negligible
if $S[\phib]>400$.  The most ``dangerous'' CCB minima, {\it i.\,e.} those
that to correspond the relatively high tunneling rates, are associated with
the third generation squarks.  This is because the action of the 
bounce is proportional \cite{chh} to the inverse Yukawa coupling squared. 

\begin{figure}
\postscript{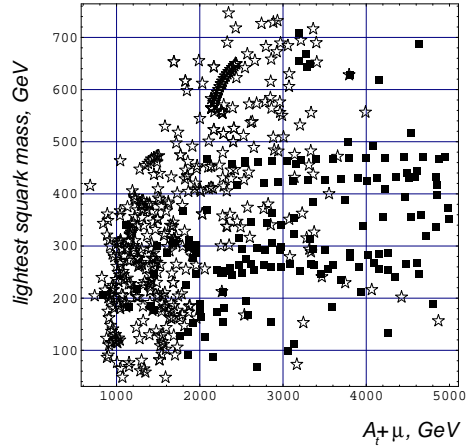}{0.3}
\caption{
The domains of stability (stars) and instability (boxes) of the
false SML vacuum with respect to tunneling into the global CCB minimum.
Light top squark and large trilinear couplings generally correspond to a
lower and thinner barrier and, thus, higher probability of tunneling.
}
\label{fig2}
\end{figure}

\begin{figure}
\postscript{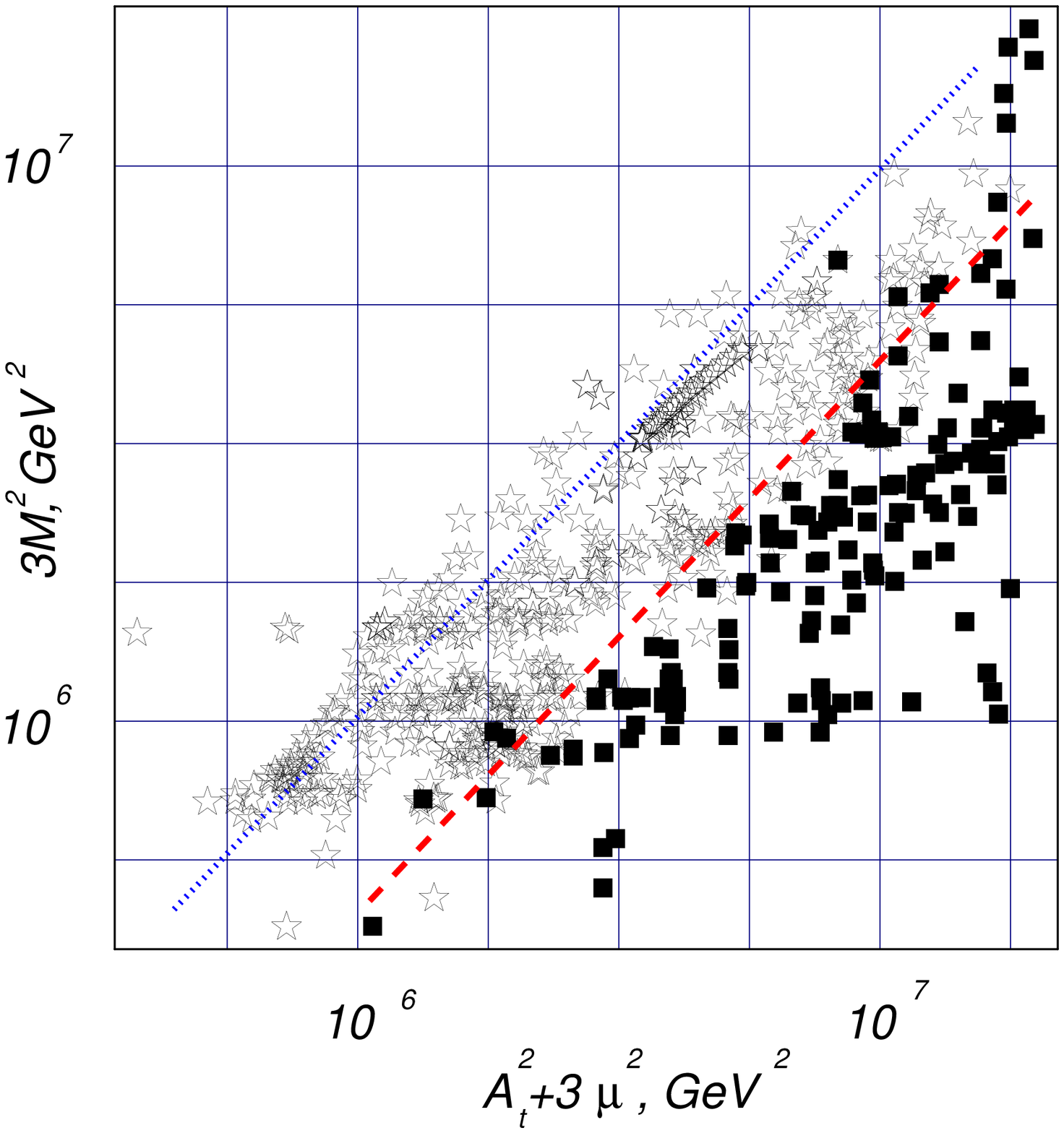}{0.3}
\caption{
The dotted line represents the empirical
criterion for the absence of the
global CCB minima: {$A_t^2+3 \mu^2 < 3 M^2$}, where
{$ M^2=m_{\tl}^2+m_{\tr}^2$}.  Taking into account the tunneling rates
relaxes this constraint to, roughly, {$A_t^2+3 \mu^2 < 7.5 M^2$}, shown as
the dashed line.  The scale is logarithmic.
}
\label{fig3}
\end{figure}

The requirement of  stability of the SML vacuum constrains the parameter
space of the MSSM \cite{kls}.    In particular, the trilinear
terms in the potential have an upper limit that depends on the quadratic 
mass terms of squarks (Fig. \ref{fig2}).  
However, these bounds are not as stringent as they
would have been, should one require the SML minimum to be the global
minimum of the potential.  We found \cite{kls} that for a large portion of 
the parameter space the presence of the global CCB minimum is irrelevant
because the time required for the Universe to relax to its lowest energy
state exceeds its present age \cite{kls}.  This is illustrated in 
Fig. \ref{fig3}. 

In summary, the color and charge conserving minimum may not be the global
minimum of the MSSM potential.  It is possible that the Universe rests in a
false vacuum whose lifetime is large in comparison to the present age of the
Universe.  Under fairly general conditions, the SML vacuum is favored by
the thermal evolution of the Universe, even if it does not represent the
global minimum.  The existence of the CCB minima of the scalar potential
results in some important constraints on models with low-energy
sypersymmetry.  However, the commonly imposed  
requirement that the SML minimum be global is too strong and may 
overconstrain the theory.

\end{document}